\documentclass[12pt]{iopart}
\usepackage{graphicx}
\begin{document}
\title[London penetration depth of Ba$\rm _{1-x}$K$\rm _x$Fe$\rm_ 2$As$\rm _2$]{London penetration depth of electron-irradiated
Ba$\rm _{1-x}$K$\rm _x$Fe$\rm_ 2$As$\rm _2$ single-crystals}
\author{P Gier{\l}owski$^1$, B Cury Camargo$^2$, I Abaloszewa$^1$, A Abaloszew$^1$, M Jaworski$^1$,  K Cho$^3$, R Prozorov$^4$, M Konczykowski$^5$}

\address{$^1$Institute of Physics, Polish Academy of Sciences, Al. Lotnik{\'o}w 32/46,\\ 02-668 Warszawa,  Poland}
\address{$^2$Faculty of Physics, University of Warsaw, ul. Pasteura 5, 02-093 Warszawa, Poland}
\address{$^3$Department of Physics, Hope College, 27 Graves Place, Holland,\\ Michigan 49423, USA}
\address{$^4$Ames Laboratory and Department of Physics and Astronomy, Iowa State University, Ames, Iowa 50011, USA}
\address{$^5$Ecole Polytechnique, CNRS-UMR 7642 and CEA/DSM/DRECAM,\\ 91128 Palaiseau, France}
\ead{Piotr.Gierlowski@ifpan.edu.pl}

\begin{abstract}
We have characterized an electron-irradiated Ba$\rm _{1-x}$K$\rm _x$Fe$\rm_ 2$As$\rm _2$ (x = 0.53) single-crystal  using two different experimental techniques: magneto-optic measurements and microwave measurements.
The crystal has been measured before as well as after the 2.5 MeV electron irradiation process. After irradiation it was annealed in a number of steps, between 90~$^\circ$C and 180~$^\circ$C, and measured after each annealing step. Most microwave measurements were performed by means of a copper cavity, taking advantage of the  TE$\rm _{011}$ and TM$\rm _{110}$ modes, allowing for the determination of the London penetration depths changes  $\delta\lambda _{ab} (T)$ and $\delta\lambda _c (T)$, ie perpendicular and parallel to the sample $c$-axis.
Appropriate equations, based on perturbation theory, were derived to calculate the penetration depths changes $\delta\lambda_{ab}$ and  $\delta\lambda_c$ for a rectangular prism geometry. The sample showed a full recovery of its  $T_c$, however the observed  behavior of $\delta\lambda _c\,$ and $\delta\lambda _{ab}\,$ was not monotonic {\it vs} annealing temperature, displaying a minimum of  $\delta\lambda _c$ and $\delta\lambda _{ab}$ at 120~$^\circ$C. This finding was confirmed by magneto-optic measurements, where besides verifying the sample uniformity and the absence of visible defects, the lower critical field $H_{c1}$ of the Ba$\rm _{1-x}$K$\rm _x$Fe$\rm_ 2$As$\rm _2$ single-crystal was obtained and the London penetration depth $\lambda _{ab}(0)\,$  was calculated.
\end{abstract}

\vspace{2pc}
\noindent{\it Keywords}: superconductivity, iron-based superconductors, electron-irradiation, microwave penetration depth, Uemura relation

\submitto{\SUST}

\maketitle

\section{Introduction}
Iron-based superconductors with a composition Ba$\rm _{1-x}$K$\rm _x$Fe$\rm_ 2$As$\rm _2$ belong to the so called '122' family of non-oxypnictides \cite{Rotter}, with a critical temperature $T_c\,$ of 38 K for $x = 0.4$. Superconductivity in these compounds is achieved by doping of barium sites with potassium, ie hole doping.
Electron-irradiation in the MeV range is known to introduce point defects (local disorder) into the crystal structure  \cite{Cho_2018}, thus reducing the $T_c\,$ of these compounds.
This work was motivated by the discrepancy of experimental data on the penetration depth of '122' iron-based
superconductors \cite{Ren,Ghigo_SUST,Li,Evtushinsky,Welp,Almoalem}, leading to a wide spread of experimental points
on the Uemura plot \cite{Uemura}. Our experimental procedure is different than the approach of other groups who
irradiated iron-based superconductors with increasing irradiation doses \cite{Cho_2016,Ghigo_Sci_Rep} because
we have used only one dose, followed by step-wise annealing procedure carried out at increasing annealing
temperatures. We were also interested to verify if the superconducting properties of an electron-beam
irradiated '122' sample can be restored by thermal annealing \cite{Marcin}. Therefore, we measured
first the magneto-optic response of the sample, followed by measurements of microwave penetration
depth changes {\it vs\,} temperature above as well as below $T_c$. Magneto-optic measurements allow
for the calculation the magnetic penetration field $\delta\lambda _{ab}$, together with the London
penetration depth $\lambda_{ab}\,$ at $T = 0$, which are combined with microwave penetration depth
changes  {\it vs\,} temperature, permitting to test the pairing-symmetry of these superconductors.

\section{Experimental}
\subsection{Sample}
We have measured a single crystal of Ba$\rm _{1-x}$K$\rm _x$Fe$\rm _2$As$\rm _2$, with x = 0.53,  grown at Ameslab   using the high-temperature FeAs flux method. Its lateral dimensions were 0.73 mm by 1.17 mm and its thickness was 20 $\mu$m. The sample was irradiated by 2.5 MeV electrons, with a dose of 11.74 C/cm$^2$, by means of a pelletron  (Sirius Accelerator at Ecole Polytechnique). After irradiation, the sample  was annealed
in air for a duration of 1 hour, starting at a temperature of 50~$^\circ$C, followed by 70~$^\circ$C and  90~$^\circ$C. Starting at 90~$^\circ$C, the temperature was increased stepwise by 10~$^\circ$C during the following annealing steps, up to 180~$^\circ$C. After each annealing process, the sample was first characterized by the magneto-optic technique.\\

\subsection{Magneto-optic measurements}
\begin{figure}[t]
    \centering
    \includegraphics[width=15cm]{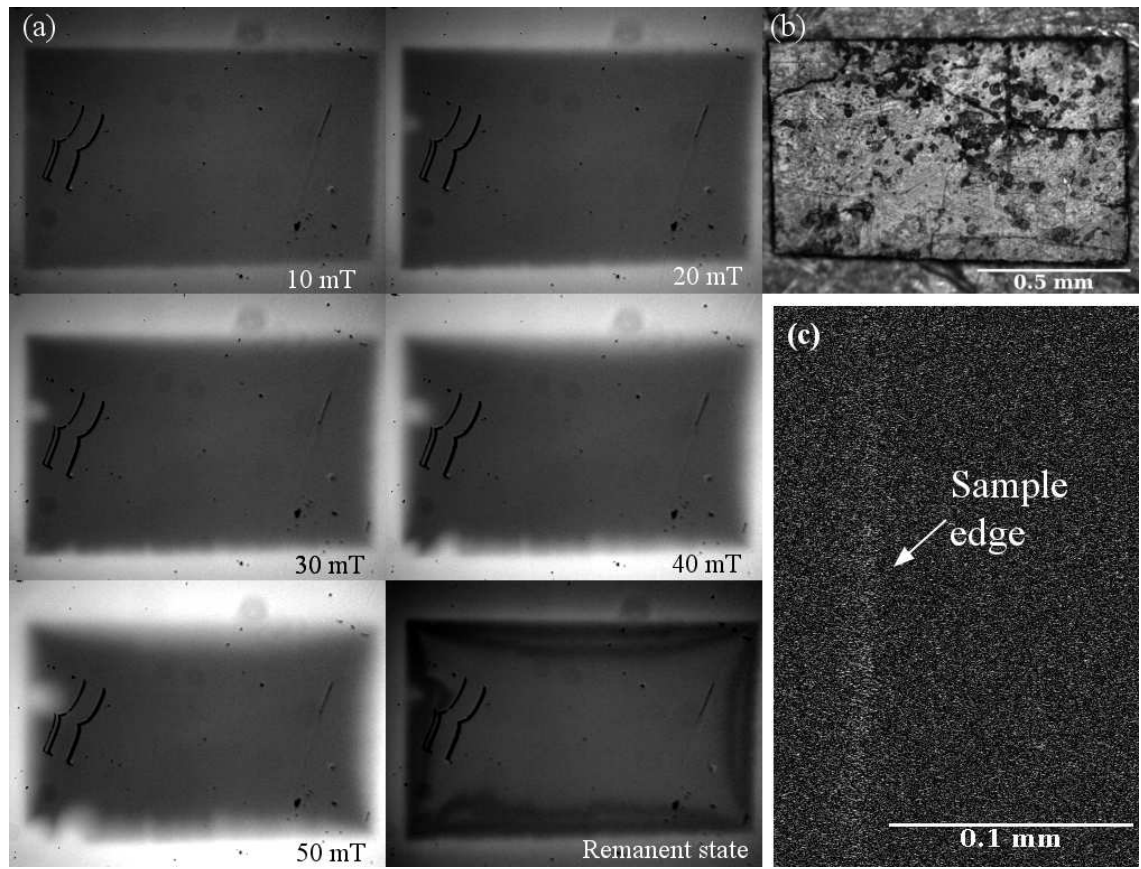}
    \caption{(a) Magnetic flux penetration and the remanent magnetic field at 10 K. (b) Sample view at room temperature. (c) Remanent state flux distribution at the edge of the sample after applying a field of 5.5 mT at 11 K.}
    \label{MO_10K}
\end{figure}

Magneto-optic (M-O) measurements were carried out in a system consisting of a continuous-flow liquid He cryostat, a polarizing
microscope, a CCD camera and a computer, in the temperature range between 4.8 and 30~K and in the magnetic field range between 0 and 500~Oe \cite{Abalosheva_2010}. As the magnetic field, directed perpendicular to the biggest surface of a superconducting sample, is slowly ramped from zero to 500 Oe, the flux penetration into the sample is visualized by a Bi-doped ferrite garnet film placed directly on the top of the sample surface, through the Faraday effect. In the Faraday effect, the rotation angle of the plane of polarization of light passing through the garnet sensor is proportional to the local magnetic field. Its effects become visible in the microscope as well as on the computer screen connected to a computer interfaced to a CCD camera mounted on top of the microscope. Figure \ref{MO_10K}(a) shows the penetration of the magnetic flux into the sample and the remanent magnetic flux distribution after the field is switched off at 10 K, visualizing the vortex matter trapped in the superconductor when it is pinned on inhomogeneities and defects in the sample. We observe that flux initially appears at the edge of the sample rather than in the center, which indicates the presence of substantial bulk pinning in the sample \cite{Okazaki_2009}.\\

\begin{figure}[t]
    \centering
    \includegraphics[width=15cm]{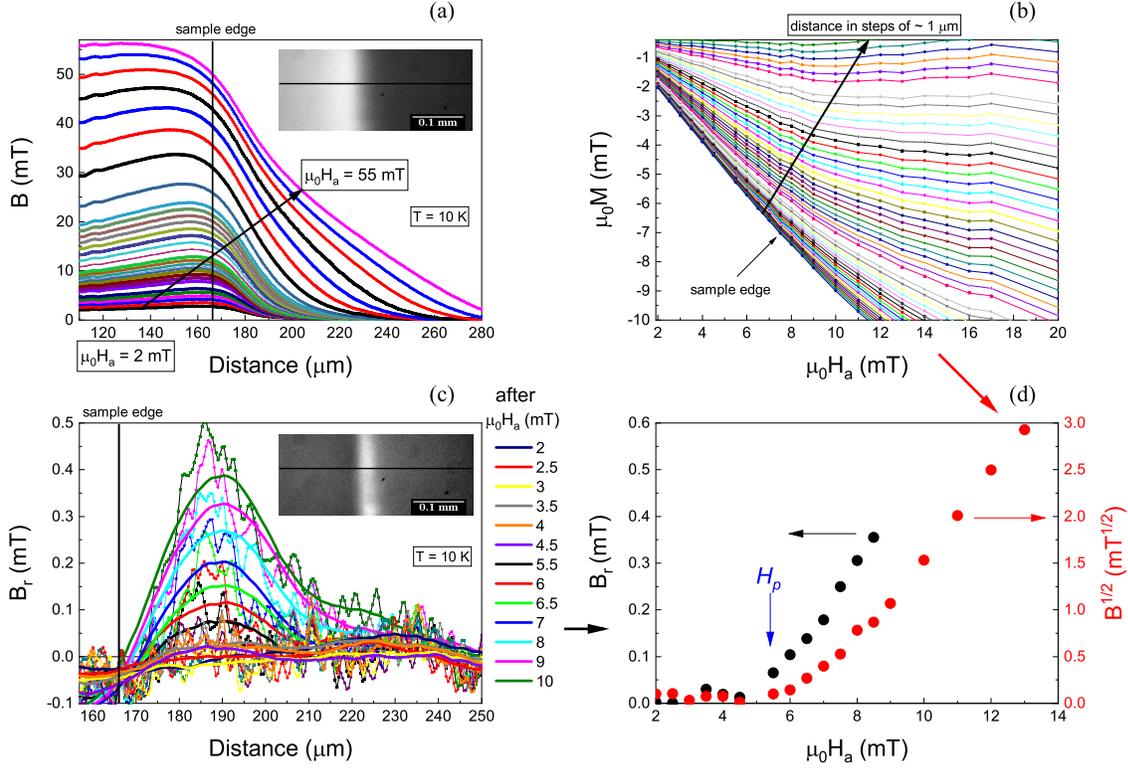}
    \caption{(a) Local magnetic induction profiles along a line perpendicular to the edge of the sample, as indicated in the M-O image fragment in the inset. Data shows measurements performed  between 2 mT and 55 mT (from bottom to top), in steps of 1 mT between 2 mT and 17 mT,
and in steps of 5 mT between 20 mT and 55 mT, at 10 K. (b) Field dependence of the local magnetization at points equally spaced ($\sim 1 \mu$m) on a line perpendicular to the edge of the sample. (c) Local magnetic induction profiles of the remanent state after the $H_a$ applying defined along the line perpendicular to the edge of the sample, thick lines correspond to smoothed data. (d) Local remanent magnetic induction $B_r$ (left axis) defined from (c) at the sample edge and $B^{1/2}=\mu_0^{1/2}(M+\alpha H_a)^{1/2}$ (right axis) defined from (b) at the sample edge plotted as function of  $H_a$. The first penetration field $H_p$ is around 5 mT.}
    \label{Hp}
\end{figure}

\noindent
The value of $H_{c1}\,$ was estimated as follows: After zero field cooling of the sample, a magnetic field was applied for a time of about 30 sec, then it was switched off and an M-O image was recorded. If the value of the applied field is sufficiently low, no remanent flux is observed. This procedure was repeated with an increasingly higher field, until the remanent flux appeared at the edge of the sample. The sample was warmed up above its critical temperature after each field sweep and then cooled down in zero field to the next desired temperature.
Figure \ref{MO_10K}(c) shows an example of the appearance of a remanent flux distribution at the sample edge.  We verified if this point in the $(H,T)$ space corresponds to the field $H_p$, at which first flux penetration occurs at the edge of the single crystal. We defined $H_p$ from the field dependence of the magnetization at the sample edge similar to the procedure described in \cite{Okazaki_2009}. In our case the planar Hall probe array was replaced by a M-O indicator, put directly on top of the sample. To determine the first penetration field, we plotted the magnetic induction profiles along a line perpendicular to the edge of the sample, as indicated in the M-O image fragment in the inset (Figure \ref{Hp}(a)). For points equally spaced on this line, we determined the field dependencies of the local magnetization $M \equiv \mu_0^{-1} B-H_a$ ($\mu _0\,$ is the magnetic permeability of the free space), which are shown in figure \ref{Hp}(b). Figure \ref{Hp}(c) shows the magnetic induction profiles of the remanent state after the applying of a certain external magnetic field $H_a$, defined also along the line perpendicular to the edge of the sample. And finally figure \ref{Hp}(d) shows a comparison of the smallest magnetic field at which the remanent state appears with the first penetration field defined as a field, above which $B^{1/2}=\mu_0^{1/2}(M+\alpha H_a)^{1/2}$ ($\alpha\,$ is the slope of the $M(H_a)\,$ initial part) starts to increase. It is clearly visible that these values are matched. $H_p\,$ was determined as a function of temperature using the acquired images of the remanent state.

\noindent
Our sample has the shape of a thin parallel prism and our $H_p\,$ measurements rely on {\it local\,} magnetic field
values and not on the {\it applied\,} external magnetic field, therefore a calculation of demagnetizing
factors was unnecessary. Further on, we did apply the approach of Polyanskii {\it et al} \cite{Polyanskii_2001}, defining
$H_{c1} \simeq H_p\,$ but being aware that $H_p\,$ is higher than $H_{c1}\,$ due
to flux pinning and the geometric barrier for flux entry close
to the sample edge. We did not apply Brandt's formula \cite{Brandt_1999} because it was
derived for a pin-free sample without magnetic flux relaxation.

\begin{figure}[t]
    \centering
    \includegraphics[width=12cm]{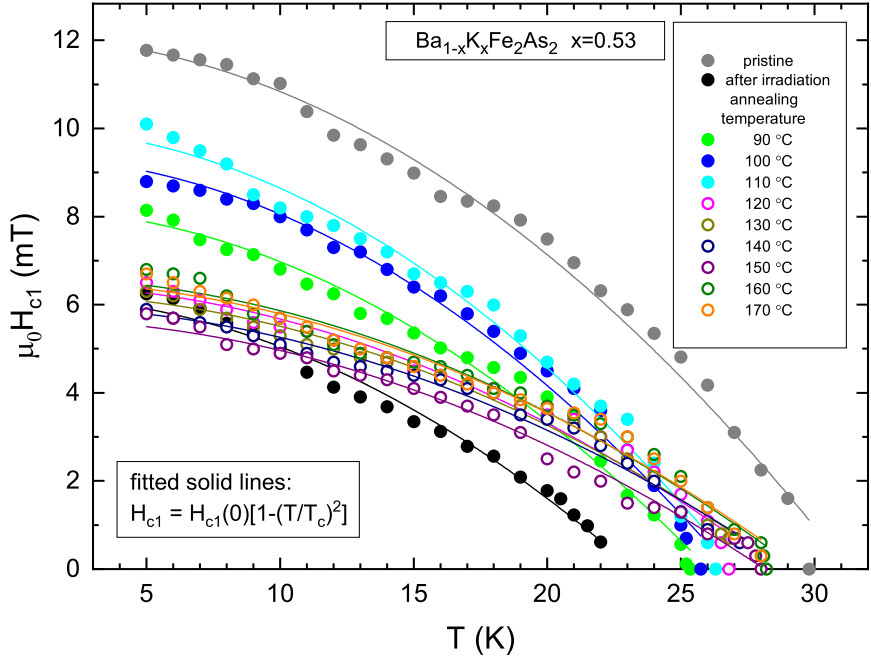}
    \caption{The critical field $H_{c1}$ {\it vs} temperature for various annealing temperatures.}
    \label{Hc1}
\end{figure}

\noindent
The experimental points shown in figure~\ref{Hc1} are accompanied by fitting lines, corresponding to the
formula $H_{c1} = H_{c1}(0)[1-(T/{T_c})^2]$ describing the temperature dependence of  $H_{c1}$ in the Meissner state of the sample, which allows one to calculate $H_{c1}$ at $T=0$.

\noindent
 The obtained curves for are not shifted parallel to each other, instead they are intersecting at certain temperatures. This finding seems at first quite unexpected because it can't be explained by a simple extended defect or crack in the measured sample (which would also manifest in the magneto-optic image). Please note also that $H_{c1}(T)$ measured for higher annealing temperatures vanishes at higher temperatures, thus indicating a steady increase of the sample $T_c$. Experimental points corresponding to $H_{c1} = 0$ were obtained at temperatures where there was no visible contrast displayed by the M-O indicator. These points were not included in the fitting procedure carried out for  $H_{c1} > 0$.

\begin{figure}[t]
    \centering
    \includegraphics[width=13cm]{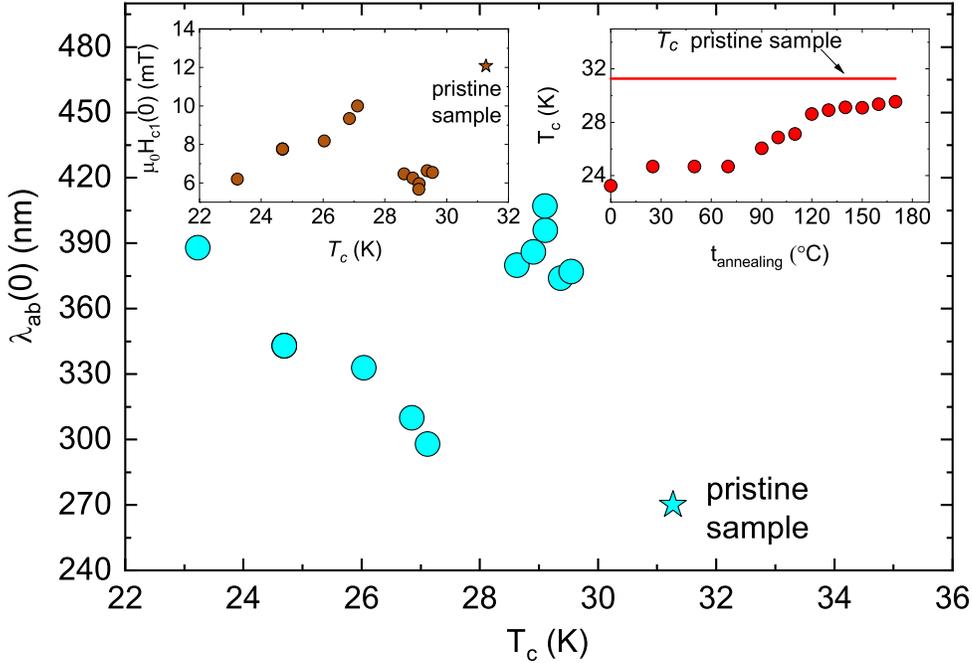}
    \caption{The penetration depths $\lambda_{ab}(0)$ for various critical temperatures ($T_c\,$ {\it vs\,} annealing temperature and $H_{c1}(0)$ {\it vs\,} $T_c\,$ are shown in the insets).}
    \label{lambda_ab}

\end{figure}

\noindent
The London penetration depth  $\lambda_{ab}\,$ in the $ab\,$ plane was calculated using the single-band formula derived by Abrikosov and Gorkov based
on the  Ginzburg-Landau theory for type II superconductors, corrected later by Hu \cite{Hu_1972}
\begin{equation}
\mu _0 H^c _{c1} ={\Phi_0\over  4\pi \lambda^2_{ab}}\left[\ln {\lambda_{ab}\over \xi_{ab}} +0.497 \right],
\end{equation}
where $\Phi_0$ is the flux quantum and $\xi_{ab}$ is the coherence length in the $ab$ plane. For our calculations, we have employed $\xi_{ab}$ values
 obtained by van der Beek {\it et al} \cite{Kees_2010}. The results are summarized in figure \ref{lambda_ab}, with $T_c$ {\it vs} annealing temperature and $H_{c1}(0\,)$  {\it vs} $T_c\,$ shown in the insets.
As can be seen, the sample annealing at temperatures of 50~$^\circ$C and 70~$^\circ$C, does not change any measured sample parameters when compared to the room-temperature point shown at
23~$^\circ$C. It can also be seen that at an annealing temperature of 120~$^\circ$C, the sample has undergone some unexpected change. Above this temperature, $T_c$ continued to rise but the $H_{c1}(0)$ decreased almost to its initial level and stopped growing, ie this change affected the vortex pinning in the sample.

\subsection{Microwave measurements}
\noindent
Most microwave measurements were performed in a cylindrical cavity made out of copper, with an inner diameter of 15 mm and the same height.
The sample prior to irradiation was measured in a superconducting niobium cavity with identical dimensions.
For the measurements, the specimen under investigation was glued on the flat end surface of a sapphire rod with a diam. of 1.9 mm, protruding 7.5 mm into the cavity from its top plate. Design details of the niobium cavity were described recently in ref \cite{Gierlowski_2019}.

\begin{figure}[t]
    \centering
    \includegraphics[width=8.5cm]{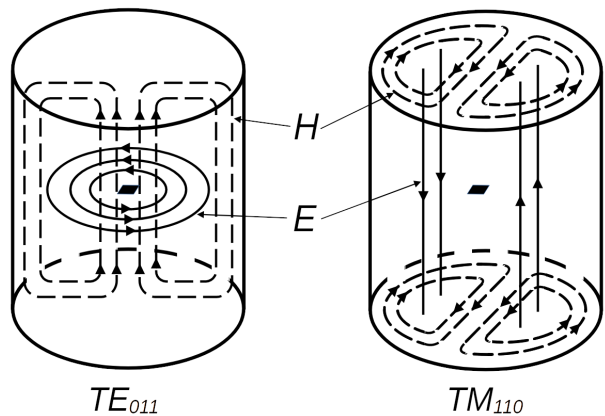}
    \caption{Electric and magnetic field lines for the TE$_{011}$ and the TM$_{110}$ modes of a cylindrical microwave cavity. The sample position is shown by a black parallelogram in the center of the cavity.}
    \label{field_lines}
\end{figure}

\noindent
In the case of the Cu cavity, we have applied two different cavity modes, namely the TE$_{011}$ and TM$_{110}$, thus
forcing the flow of microwave currents in our sample in different manners: along the $ab$-plane of the single-crystal for the former and
both perpendicular as well as along the $ab$-plane for the latter.
However, the circulation of microwave currents across the junctions between the cylindrical part and the endplates of the cavity, generated by the TM$_{110}$ mode, causes its unloaded quality factor Q$_0\,$  to be 4 times smaller than Q$_0$  of the TE$_{011}$ mode. In the latter, wall currents do not intersect these junctions, thus resulting in smaller losses.
The frequencies for the TE$_{011}$ and TM$_{110}$-modes were around 26.2 GHz and 23.9 GHz, respectively. A simple plot showing their electric and magnetic field lines  \cite{Gandhi} is presented in figure \ref{field_lines}. Their excitation could be tuned by controlling the angle of the loop-antenna  \cite{Gierlowski_2019}.

\begin{figure}[t]
    \centering
    \includegraphics[width=11.5cm]{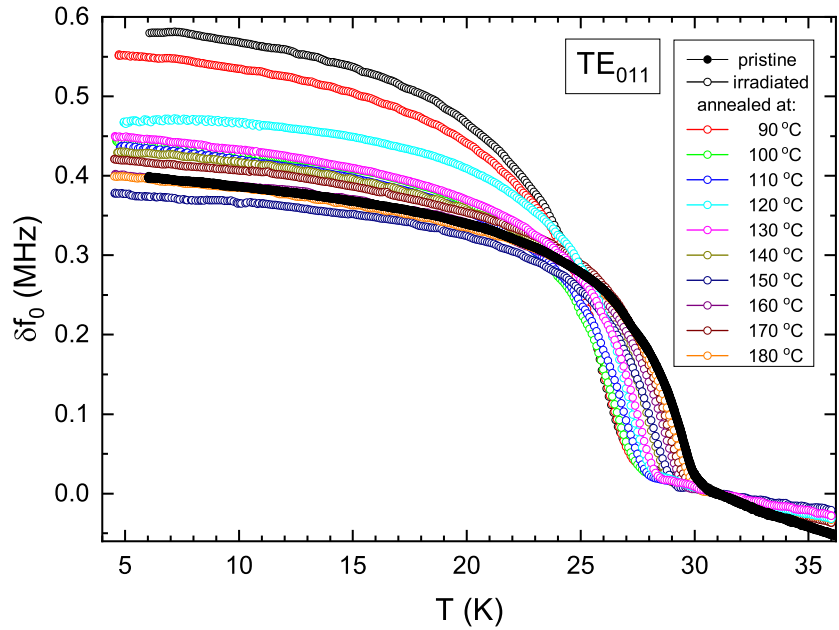}
    \caption{The frequency change of the TE$_{011}$ resonance for various annealing temperatures of the Ba$\rm _{1-x}$K$\rm _x$Fe$\rm_ 2$As$\rm _2$ sample, normalized at 31~K.}
    \label{f0_vs_T}
\end{figure}

\begin{figure}[b]
    \centering
    \includegraphics[width=11.5cm]{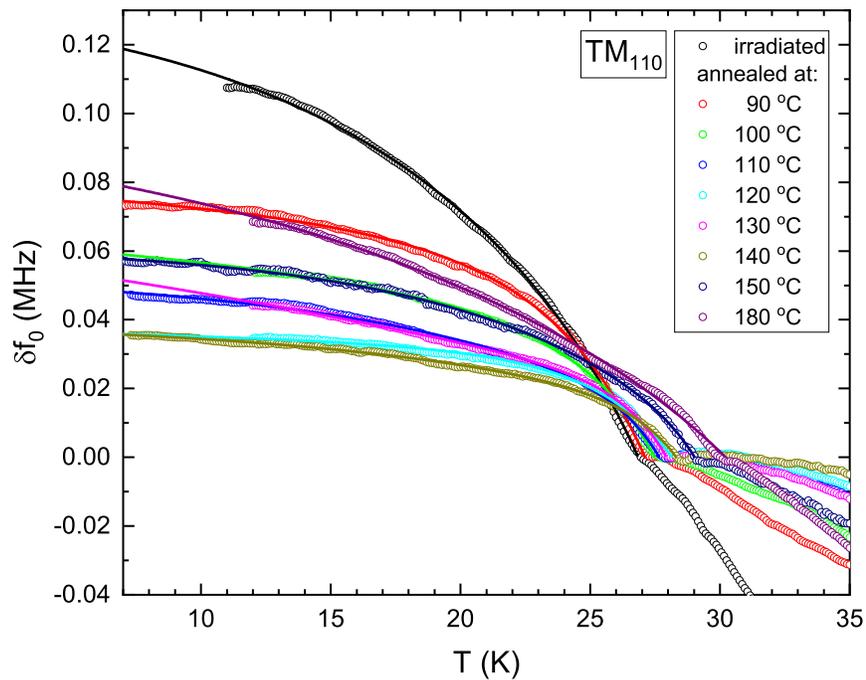}
    \caption{The frequency change of the TM$_{110}$ resonance for various annealing temperatures of the Ba$\rm _{1-x}$K$\rm _x$Fe$\rm_ 2$As$\rm _2$ sample, set to zero at the $T_c\,$ of the sample. Solid lines of the same color are fitting lines below $T_c$ of the data given by equation \ref{EXPDEC3}.}
    \label{delta_f0_TM}
\end{figure}

\noindent
The microwave measurements were preceded by a calibration procedure of the semirigid coaxial cable connecting the cavity to the Vector Network Analyzer (VNA). During the measurement, we have set
the microwave power level to -10 dBm, in order to avoid excessive sample heating. The frequency sweep was set to 5 MHz around the resonance frequency.
In addition, before each measurement {\it vs\,} temperature, a sweep of $f_0$ and $f_L$ {\it vs\,} the phase offset angle was performed and it was set to the position where the unloaded resonance frequency $f_0$ is equal to the loaded resonance frequency $f_L$. This was done in order to avoid artifacts caused by the small length (a few millimeters)  of the semirigid cable piece between the microwave connector attached directly to the cavity housing and the loop-antenna \cite{Canos_2006} as this part was not included in the calibration procedure.

\noindent
During a frequency sweep recorded at each temperature, typically 2001 data points were measured. A Smith circle was fitted using 401 points close to resonance center frequency and the loaded and unloaded Q-factors, $Q_0\,$ and $Q_L$, respectively, were calculated based on the Kajfez  procedure \cite{Kajfez_book}. The  resonance frequencies $f_L\,$ and  $f_0$ were calculated using the same method.

\noindent
After subtracting the frequency change for the empty cavity measured {\it vs\,} $T$, we obtained the plots presented in figures \ref{f0_vs_T} and \ref{delta_f0_TM}.
However, the observed frequency shift is not monotonic
{\it vs\,} annealing temperature. The data presented in figure \ref{f0_vs_T}  will be used later for calculating $\delta\lambda_{ab}$. In order to be able to calculate also $\delta\lambda_c$,  results presented in figure \ref{delta_f0_TM} will be used.

\noindent
Solid lines in figure \ref{delta_f0_TM} are fitting lines described by
the following equation

\begin{equation}
f_0 = f_{constant} + \sum _{i=1} ^{i=3} a_i \exp ( {-T \over b_i} ),
\label{EXPDEC3}
\end{equation}
\noindent
where  $f_{constant}$, $a_i$ and $b_i$ and are constants, different for each annealing temperature.

\noindent
During the measurement of the TM$_{110}$ mode, we observed two additional phenomena occurring at the same time,
accompanying the frequency change of this mode associated with the superconducting transition. The first one
was a small slow drift of the resonance frequency at a constant rate during the temperature sweep. The second one
was a notable hysteresis of the measured resonant frequency data {\it vs\,} temperature  when measuring a full temperature
loop during the sweep. The effects of both phenomena were taken into account before further processing our data by
first removing the linear frequency drift, followed by averaging the measured hysteresis loops which were most likely caused
by  thermal properties of our 3D-printed plastic piece used for the support of the sapphire rod.
The results of the TM$_{110}$ data after getting rid of these two factors are shown in figure \ref{delta_f0_TM}.
However, we did not observe a frequency drift, nor a hysteretic behavior, for the TE$_{011}$ mode, despite the same plastic piece
being used to hold the sapphire rod.

\subsection{Penetration depth: experiment}

\noindent
\subsubsection{Perpendicular orientation}
\begin{figure}[h]
    \centering
    \includegraphics[width=12.5cm]{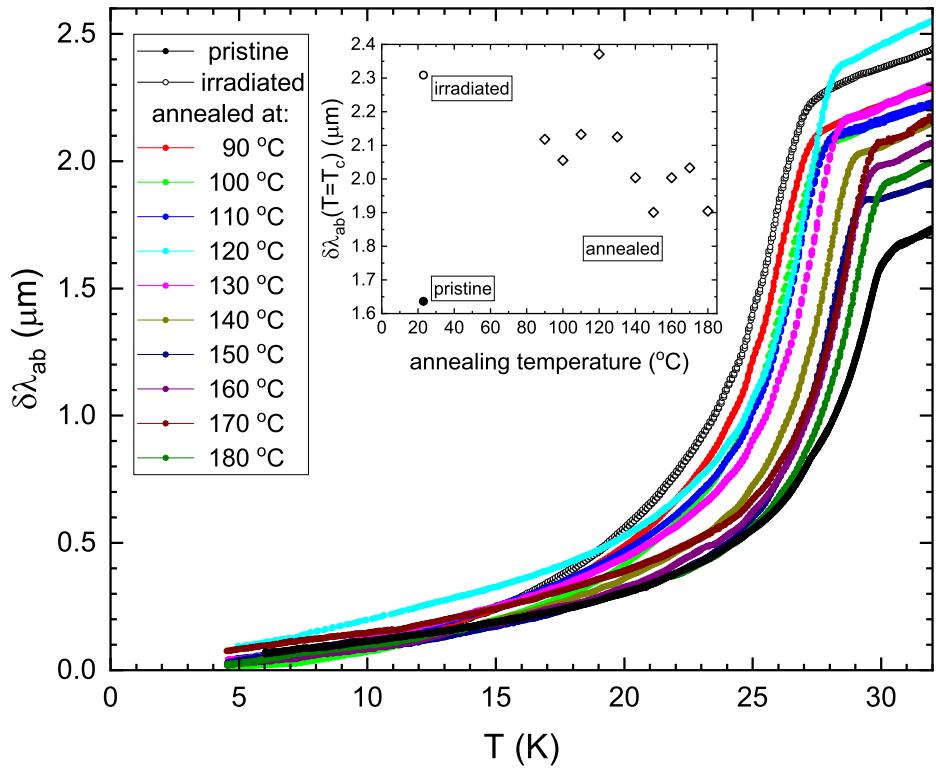}
    \caption{The variation of the London penetration depth in the ab-plane after various annealing temperatures of the Ba$\rm _{1-x}$K$\rm _x$Fe$\rm_ 2$As$\rm _2$ sample, measured for the TE$_{011}$-mode. The penetration depth $\delta\lambda_{ab}$ at $T_c\,$ is shown in the inset.}
    \label{delta_lambda}
\end{figure}

The procedure how to treat experimental data in order to obtain penetration depth values is described in the Appendix.
For the perpendicular orientation we chose $f_e(0)$=26.236272 GHz with the prefactors $\Gamma_\bot \,$ of  $1.312\cdot10^{-3}$, $1.019\cdot10^{-3}$,  and $0.992\cdot10^{-3}$.
 The first $\Gamma_\bot \,$ was applied for the untreated sample, the irradiated sample and the one annealed at 90~$^\circ$C.
 The second $\Gamma_\bot \,$ was applied for the one annealed between 100~$^\circ$C and 140$^\circ$C, and the third prefactor $\Gamma_\bot \,$ for the one annealed between 150~$^\circ$C and 180~$^\circ$C.
The reason for using three values of the prefactors $\Gamma_\bot \,$ was the mechanical instability of our heavily electron-irradiated sample, ie its size decreased slightly during the measurement sequence, due to unwanted and unexpected breaking away of very small crystalline pieces from the main body of the remaining crystal. The obtained results are presented in figure \ref{delta_lambda}.

\begin{figure}[h]
    \centering
    \includegraphics[width=12.5cm]{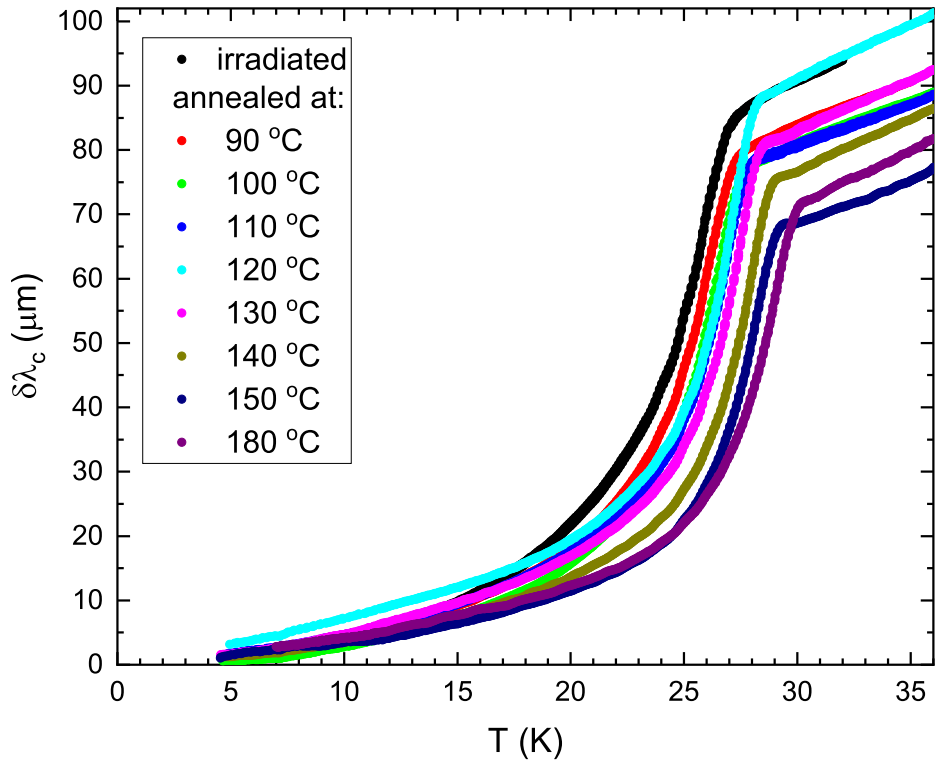}
    \caption{The variation of the London penetration depth in the $c$-axis direction after various annealing temperatures of the Ba$\rm _{1-x}$K$\rm _x$Fe$\rm_ 2$As$\rm _2$ sample, calculated based on the TE$_{011}$ and TM$_{110}$-modes.}
    \label{delta_lambda_c_TM}
\end{figure}

\subsubsection{Parallel orientation}
\noindent
Substituting the sample dimensions  into Eq. (\ref{pert2a}) and next into Eq. (\ref{pert3a})
we find, for our cavity dimensions  $R=7.5$ mm and  $L=15$ mm, the prefactor $\Gamma_\|\,$ corresponding to the parallel orientation
\begin{equation}
\label{gcalc}
\Gamma_\|=abc\cdot 0.5814\times 10^{-3},
\end{equation}
where $a$,  $b$ and $c\,$ are the sample dimensions in mm.

\subsection{Discussion}
\begin{figure}[t]
    \centering
    \includegraphics[width=12cm]{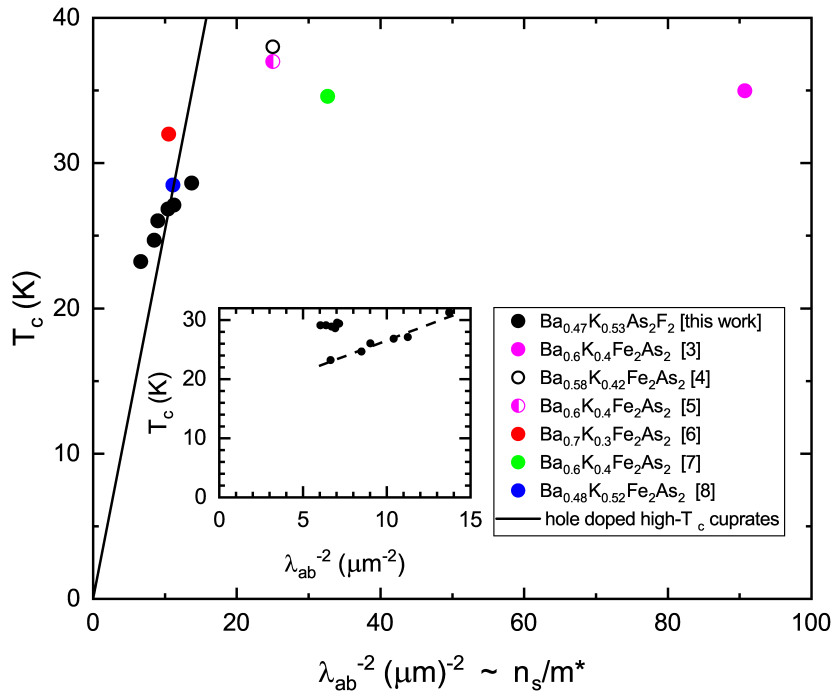}
    \caption{The Uemura plot for our sample together with data for other Ba$\rm _{1-x}$K$\rm _x$Fe$\rm_ 2$As$\rm _2$ samples.
    The solid line corresponds to the Uemura plot for hole doped cuprates.
    The dashed line in the inset is a linear fit of the experimental points. }
    \label{Uemura}

\end{figure}

\noindent
The experimental Uemura-relation \cite{Uemura}, namely $T_c\; \sim 1/{\lambda^2_{ab}}\; \sim n_s/m^{\ast}$, where $n_s\,$ is the superfluid density and $m^{\ast}\,$ the effective mass, was initially discovered for high-$T_c\;$ superconductors based on copper oxides. It is unclear, whether this  relation also holds for iron-based superconductors.
The Uemura-plot for our sample, based on the results of M-O measurements for annealing temperatures below 120~$^\circ$C, is shown in figure \ref{Uemura}, together with  data of several other Ba$\rm _{1-x}$K$\rm _x$Fe$\rm_ 2$As$\rm _2$ samples with different x values, extracted from the literature \cite{Ren,Ghigo_SUST,Li,Evtushinsky,Welp,Almoalem}. The inset shows all our points. Note that all our points below the annealing temperature of 120~$^\circ$C do follow approximately a dotted straight line, shown in the inset, whereas points corresponding to higher annealing temperatures do not. However, this line does not pass through the origin. This indicates that the experimental situation is more complex, most likely due to multiple electron bands.
We also did not observe a saturation or decrease of $T_c$ {\it vs} $\lambda_{ab}^{-2}$, as is the case for YBCO, La$\rm_{2-x}$Sr$\rm_x$CuO, Bi-2212, Bi-2223, and Pr-doped YBCO \cite{Uemura,Seaman_1990}. The behavior of our sample is in the vicinity of the Uemura relation for hole-doped cuprates.

\noindent
\\

\begin{figure}[t]
	\centering
	\includegraphics[width=10cm]{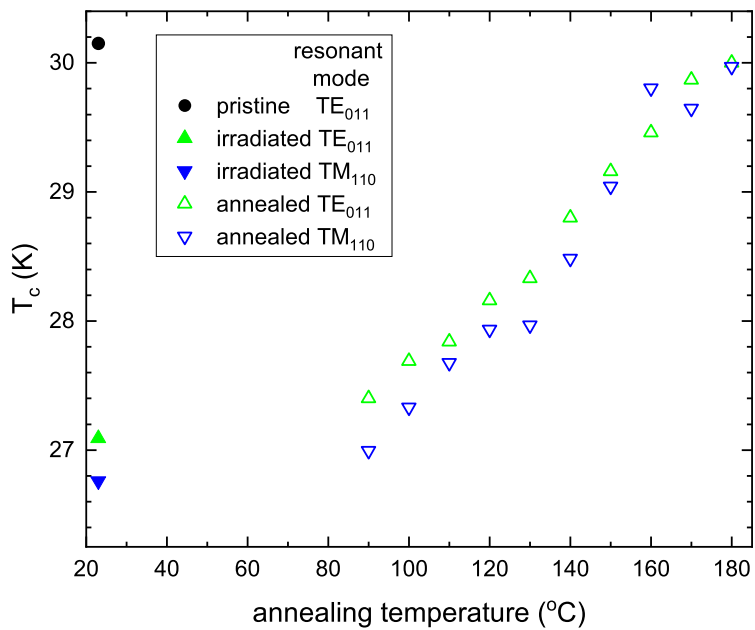}
	\caption{The critical temperature $T_c$ after various annealing temperatures of the Ba$\rm _{1-x}$K$\rm _x$Fe$\rm_ 2$As$\rm _2$ sample, for the
TE$_{011}$ and TM$_{110}$ modes.}
	\label{Tc_vs_annealing_temp}
\end{figure}

\noindent
 The critical temperatures were extracted from $f_0\,$ {\it vs\,} $T\,$ measurements as the intersection of linear extrapolations
 of $f_0(T)\,$ above and below $T_c$.
The results of this procedure are shown in figure \ref{Tc_vs_annealing_temp} for both TE$_{011}$ and TM$_{110}$ modes.
The  points drawn at 23~$^o$C (room temperature) correspond to the $T_c\,$ of the pristine sample and to the irradiated sample before annealing at higher temperature. $T_c\,$ increases monotonically with annealing temperature and reaches almost its initial value before electron irradiation. The average increase of $T_c\,$ {\it vs\,} annealing temperature is 0.326 K per 10~$^o$C.

\noindent
The $T_c\,$ value for the irradiated sample measured by the M-O technique was around 22.5~K, which is about 4.5~K below the temperature obtained by microwave methods. This difference can be explained by a self-annealing process occurring during the period of 23 weeks between these measurements, during which the sample was stored at room temperature. After sample annealing at 90~$^\circ$C, our measurement sequence was accelerated, leading to a reduction of $T_c\,$ differences observed by the two methods after the subsequent annealing steps.

\noindent
The  penetration depth variation measured at $T_c$, $\delta\lambda_{ab} = 1.7 \mu$m  (called the skin depth in the $ab\,$ plane), is similar to the results of Ghigo {\it et al\,} \cite{Ghigo_IEEE}, even though both samples are not identical.
In the case of Ghigo {\it et al}, the K content x in Ba$\rm _{1-x}$K$\rm _x$Fe$\rm_ 2$As$\rm _2$ was lower (x=0.42),  whereas in our case x=0.53.
As expected, the electron irradiation increased the penetration depth which can be explained by introducing structural disorder in the form of Frenkel pairs \cite{Cho_2018}.
Please note that the penetration depth did not recover completely its value prior irradiation after annealing, contrary to the critical temperature $T_c\,$.

\begin{figure}[t]
	\centering
	\includegraphics[width=9cm]{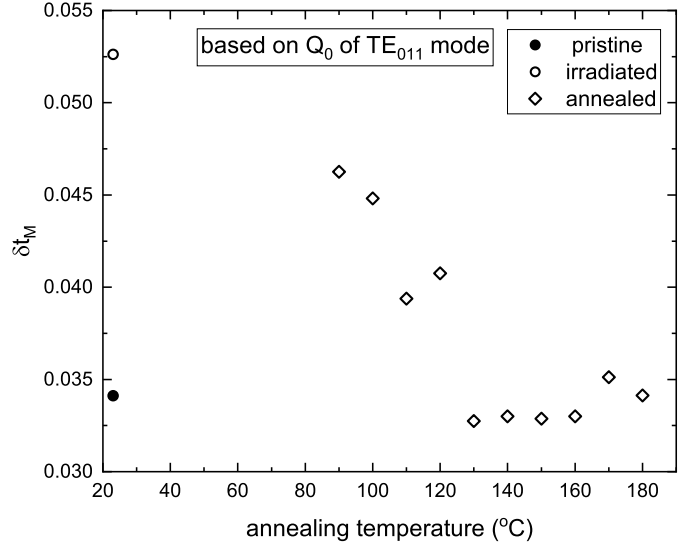}
	\caption{The normalized transition width $\delta t_M\,$ for various annealing temperatures of the Ba$\rm _{1-x}$K$\rm _x$Fe$\rm_ 2$As$\rm _2$ sample, measured for the TE$_{011}$-mode.}
	\label{Tm}
\end{figure}

\noindent
Another parameter characterizing a superconductor is the width of its transition to the superconducting state. It was impossible to attach contacts for resistivity measurements to our sample because of potential disruptions to the M-O and microwave measurements. Therefore we have introduced and calculated a parameter $\delta t_M$ corresponding to the transition width as follows.
First, the $T_c\,$ was found based on $Q_0$-factor measurements {\it vs} temperature but its width could not be immediately calculated based on these measurements because of the slow change of $Q_0$ {\it vs} $T$. Therefore, we have normalized the $Q_0(T)\,$ plots for the pristine, irradiated and annealed samples, respectively, by converting them to  $Q_0(t)\,$ plots, where $t\,$ is the reduced temperature ($t = T/T_c$\,) and calculated a 9th degree polynomial fit for the discrete data points close to $T_c\,$, in the range $0.8 < t < 1.05$. These fits allow one to calculate their derivatives and to find the temperature  $t_M\,$ where $|{dQ_0}/{dt}|$ reaches its maximum, corresponding to the maximum slope of $Q_0(t)$. After subtracting $1-t_M$, we obtain a number $\delta t_M\,$ describing the normalized 'transition width' defined above. The results of this procedure are shown in figure \ref{Tm} where a recovery of the initial transition width $\delta t_M\,$ after annealing is clearly visible.

\noindent
We have also approximated the $\delta\lambda_{ab}$ {\it vs} temperature behavior using a simple power-function of the following form
\begin{equation}
\delta\lambda_{ab}(t) = a + bt^c,
\end{equation}
where $t$ is the reduced temperature, $a\,$ and $b\,$ are constants calculated during the fitting procedure, and $c\,$ is an exponent, typically in the range between 1 and 6.5.\\

\begin{figure}[t]
    \centering
    \includegraphics[width=9cm]{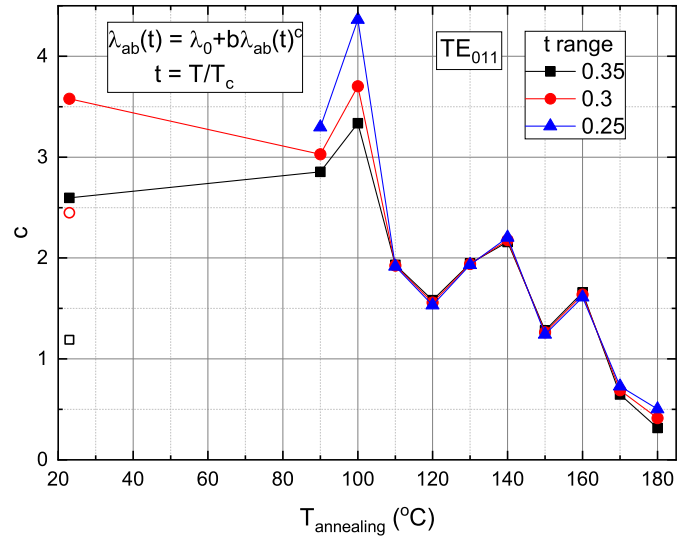}
    \caption{The exponent c {\it vs} annealing temperatures of  Ba$\rm _{1-x}$K$\rm _x$Fe$\rm_ 2$As$\rm _2$ sample, before (open symbols) and after electron-beam irradiation (closed symbols).}
    \label{exponent_c}
\end{figure}

\noindent
A plot of this exponent {\it vs\,} annealing temperature is shown in figure \ref{exponent_c}, for a few different maximum ranges of $t\,$.
The initial exponent values (for the pristine sample) are displayed as open symbols. After the electron-beam irradiation, an increase of these exponents becomes clearly visible. Their values do depend very little on the temperature range of the data points in some areas of the annealing temperature, especially above the annealing temperature of 110~$^o$C.
Our experimental data look slightly inconsistent with the results of Cho {\it et al\,} \cite{Cho_2016}.
On one hand, the $T_c\,$ of our pristine sample is around 31~K, whereas one would expect 33~K.
On the other hand, the $n$-value increases after irradiation, which is consistent with the results
of \cite{Cho_2016}. However, after annealing at higher temperature it drops below 1, whereas a number above 1 would be expected -
corresponding to the pristine state.
The decrease of the exponent $c\,$ could indicate a continuous change of the properties of the sample,
from a fully gapped s-wave superconductor after electron irradiation, to a clean d-wave superconductor, according to the paper of Ghigo {\it et al\,}\cite{Ghigo_Sci_Rep}. In their case, however, the  sample was irradiated by 250 MeV gold ions and not by electrons.

\noindent
As can be noticed, the $\delta\lambda_c\,$ results shown in figure \ref{delta_lambda_c_TM} exceed the sample thickness but not the lateral sample dimensions. Thus, it is not necessary to modify the formulae applied for these calculations.
As can be easily seen, the results obtained for the sample using two different experimental techniques are in good agreement, although
some quantities derived, like $T_c\,$, present slight differences.

\noindent
The next step was the calculation of the anisotropy ratio $\rho(T)$ of our sample, defined as
\begin{equation}
\label{rho}
\rho (T) = {\delta\lambda_c}(T)/{\delta\lambda_{ab}(T)},
\end{equation}
where $\delta\lambda_{ab}(0) =\delta\lambda_c(0) = 0$.\\

\begin{figure}[h]
    \centering
    \includegraphics[width=10cm]{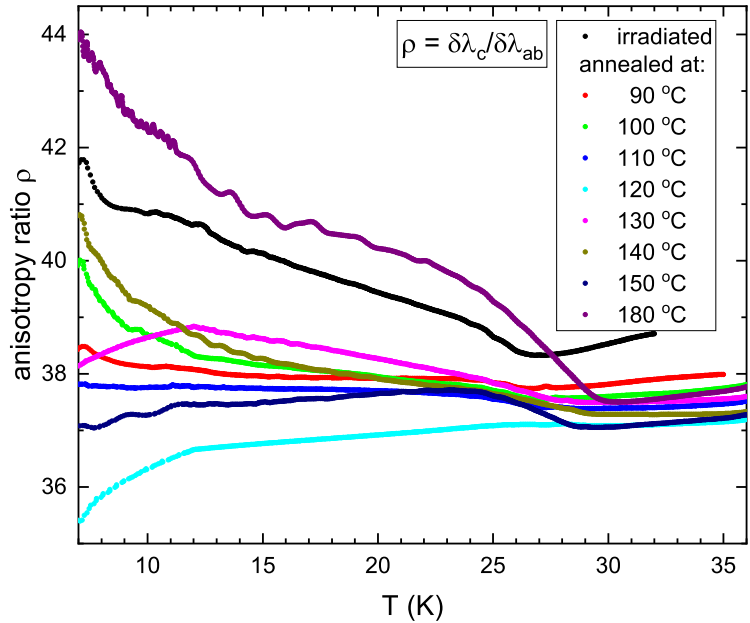}
    \caption{The anisotropy ratio $\rho $ after various annealing temperatures of the Ba$\rm _{1-x}$K$\rm _x$Fe$\rm_ 2$As$\rm _2$ sample.}
    \label{anisotropy}
\end{figure}

\noindent
The results are presented in figure \ref{anisotropy}, where $\rho\,$ is shown in the temperature range between 7~K and 35~K.
In general, the magnitude of  $\rho\,$ was in the range between 35 and 44, presenting a decrease for the sample annealed between 90 and 120~$^\circ$C for all temperatures below $T_c\,$, followed by an increase for treatments at higher temperature. Above $T_c\,$  only small differences of $\rho\,$ are visible, which occur solely for samples annealed above 100~$^\circ$C. Thus, in our case the magnitude of $\rho\,$ is much higher than the anisotropy ratio between 3 to 7 typically found for unirradiated Ba$\rm _{1-x}$K$\rm _x$Fe$\rm_ 2$As$\rm _2$ superconductors \cite{Prozorov_2009}, and closer to the anisotropy ratio of '1111' iron-based superconductors \cite{Martin_2009}.

\section{Conclusions}
Through M-O and microwave measurements, we observed an almost-complete recovery of the superconducting properties of 2.5 MeV e-beam irradiated single-crystalline Ba$\rm _{1-x}$K$\rm _x$Fe$\rm_ 2$As$\rm _2$ (x = 0.53), by subjecting the material to stepwise annealing up to 180~$^o$C. An anomaly of the sample properties, confirmed by both techniques, was found after
annealing at 120~$^o$C. The obtained results for $\lambda_{ab}\,$ are very close to the Uemura line observed for hole-doped cuprate superconductors,
although the straight fitting line does not pass through the origin.
Due to the much smaller change of the resonant frequency for the TM$_{110}$ mode, as compared to the TE$_{011}$ mode,
the measurement accuracy is reduced in the former case. Equations based on cavity perturbation theory for the calculation of the penetration depths $\delta\lambda_c\,$ and  $\delta\lambda_{ab}\,$, using a uniform approach for both modes, were derived for the case of a parallel prism shaped sample.
The  exponent $c\,$ associated with the power-law behavior of $\delta\lambda_{ab}\,$  {\it vs} annealing temperature is typical for the transition from a fully gapped clean {\it s\,}-wave to a clean {\it d\,}-wave superconductor.

\ack
This work was supported in part by the National Science Center, Poland, research project no 2014/15/B/ST3/03889.
We thank Y Liu and T A Lograsso for providing the single crystal of Ba$\rm _{1-x}$K$\rm _x$Fe$\rm_ 2$As$\rm _2$ (x = 0.53). The work in Ames Laboratory was supported by the US Department of Energy (DOE), Office of Science, Basic Energy Sciences, Materials Science and Engineering Division. Ames Laboratory is operated for the US DOE by Iowa State University under contract DE-AC02-07CH11358.

\begin{appendix}
\setcounter{section}{1}
\section*{Appendix: Penetration depth calculations: theory}
\subsection*{Introduction}
\noindent
No analytic solutions of the electric and magnetic field components exist for small samples having the
shape of a rectangular prism with finite dimensions, placed inside a cylindrical microwave cavity.
However, exact solutions were found in the case of a sample having the form of a oblate ellipsoidal body.
Thus, taking advantage of the latter, one can approximate the rectangular prism shape by a corresponding ellipsoidal one,
keeping in mind the limited accuracy of this approach which will be discussed later.\\

\noindent
Based on the data shown above, the London penetration depths were calculated using the procedure described below.
We can take advantage of the properties of two cavity-modes, namely TE$_{011}$ and TM$_{110}$.
For the TE$_{011}$ cylindrical mode, a locally uniform magnetic field is directed {\it perpendicular\,} to the sample plane and for the TM$_{110}$ cylindrical mode the locally uniform magnetic field is {\it parallel\,} to the sample plane.

\noindent
For the parallel orientation the magnetic field around the sample is practically
undistorted. As a result, the penetration depth can be easily determined using
e.g. a simple approach based on the energy balance \cite{Trunin_2005}. However,
for the perpendicular orientation the magnetic field differs considerably from
its initial uniform distribution, and a proper evaluation of the geometric
factor (responsible for the field distribution around the sample) is crucial
for the reliable determination of the penetration depth.

\noindent
To evaluate the geometric factor experimentally some authors \cite{Trunin_2005,
Mao} suggest to measure test samples of the same geometry and known
parameters. Another possibility is to compare the dc resistivity of the
sample with its normal state surface impedance \cite{Mao}. On the other hand,
some approximate theoretical results have been published \cite{Trunin_2005,
Trunin_1998} making it possible to evaluate the geometric factor e.g. for
square plates ($a=b\gg c$) or long thin needles ($b\gg a,c$).

\noindent
Contrary to various theoretical methods reported so far in the literature,
in this paper we present a unified approach using the same formalism for
the analysis of both perpendicular and parallel orientation. The results
are rigorous for an oblate ellipsoid but appear to be fairly accurate
for a flat prism-like rectangular sample.

\subsection{Theoretical background}
\noindent
We start with the general perturbation formula for a microwave cavity
\cite{Harr}
\begin{equation}
\label{pert1}
\frac{\omega-\omega_0}{\omega_0}=-\frac{\int(\Delta\epsilon EE^\ast_0+
\Delta\mu HH^\ast_0)dv}{\int(\epsilon_0 EE^\ast_0+\mu_0 HH^\ast_0)dv},
\end{equation}
where $\omega_0=2\pi f_0$,
$E_0$ and $H_0$ denote the angular resonant frequency,
the electric and magnetic field distribution for an
unperturbed cavity, whereas $\omega=2\pi f$, $E$ and $H$ denote the
corresponding quantities for a perturbed cavity and both integrals are
calculated for the entire cavity volume.

\noindent
$\Delta\epsilon=\epsilon_0(\epsilon_r-1)$, $\Delta\mu=\mu_0(\mu_r-1)$ and
$\epsilon_r$, $\mu_r$ are the relative permittivity and permeability
of the perturbing object.

\noindent
If the perturbing object (sample) is small, then the denominator of
Eq. (\ref{pert1}) can be approximated as
\begin{equation}
\label{int}
\int\epsilon_0 EE^\ast_0 dv \simeq \int\mu_0 HH^\ast_0 dv \simeq
\int\epsilon_0 |E_0|^2 dv = \int\mu_0 |H_0|^2 dv
\end{equation}
In particular, for a magnetic perturbation we have $\Delta\epsilon=0$ and
\begin{equation}
\label{pert2}
\frac{f-f_0}{f_0}=-\frac{(\mu_r-1)\int_{V_s}HH^\ast_0 dv}
{2\int_{V_c}|H_0|^2 dv},
\end{equation}
where $V_s$ and $V_c$ denote the sample and cavity volume, respectively.

\noindent
For a special case of a hypothetic ellipsoidal sample placed in a locally
uniform magnetic field, the perturbed field $H$ within a sample is also
uniform and given by \cite{Harr}
\begin{equation}
\label{hint}
H=\frac{H_0}{1+N(\mu_r-1)},
\end{equation}
where $N$ denotes the demagnetization factor \cite{Osb}.

\noindent
For a superconducting sample we can take $\mu_r=0$, thus
substituting Eq.~(\ref{hint}) into Eq. (\ref{pert2}) we obtain
\begin{equation}
\label{pert4}
\frac{f-f_0}{f_0}=\frac{1}{1-N}\cdot
\frac{\int_{V_s}|H_0|^2 dv}{2\int_{V_c}|H_0|^2 dv}
\simeq \frac{1}{1-N}\cdot \frac{|H_{\rm max}|^2 V_s}
{2\int_{V_c}|H_0|^2 dv},
\end{equation}
where $\int_{V_s}|H_0|^2 dv\simeq |H_{\rm max}|^2 V_s$,
and $H_{\rm max}$ denotes the maximum value of the magnetic field
in the cavity center where the sample is placed.
\noindent

\subsubsection{Perpendicular orientation}

\noindent
We consider now a flat ellipsoidal sample in the perpendicular
orientation, i.e. with the magnetic field $H_{\rm max}$ directed
perpendicularly to the flat face of a sample. For $\bar{a}>\bar{b}\gg\bar{c}$
($\bar{a}$, $\bar{b}$, $\bar{c}$ being semi-axes of an ellipsoid) we find
\cite{Osb}
\begin{equation}
\label{n}
1-N=(\bar{c}/\bar{b})E(k,\theta),
\end{equation}
where
$k=\sin\phi/\sin\theta$, $\cos\phi=\bar{b}/\bar{a}$,
$\cos\theta=\bar{c}/\bar{a}$, and $E(k,\theta)$ is the elliptic
integral of the second kind. For a flat sample $\bar{c}\ll\bar{a}$, thus
$k\simeq[1-(\bar{b}/\bar{a})^2]^{1/2}$, $\theta\simeq\pi/2$, and $E(k,\theta)$
can be replaced by the complete elliptic integral of the second kind
${\rm\bf{E}}(k)$.

Substituting Eq. (\ref{n}) into Eq. (\ref{pert4}) and using
$V_s=(4/3)\pi\bar{a}\bar{b}\bar{c}$ we obtain
\begin{equation}
\label{pert5}
\frac{f-f_0}{f_0}=\frac{(4/3)\pi\bar{a}\bar{b}^2}
{{\rm\bf{E}}(k)}\cdot
\frac{|H_{\rm max}|^2}{2\int_{V_c}|H_0|^2 dv}.
\end{equation}

For real samples (such as e.g. discs or thin rectangular prisms) the simple
approach presented above cannot be used directly.
Nevertheless, the demagnetization factor $N$ is
still a useful tool in the analysis of more realistic sample shapes
if we consider $N$ in Eq. (\ref{pert4}) as an effective
(averaged) quantity $N_{\rm eff}$ \cite{Proz1,Proz2}.
Moreover, $N_{\rm eff}$ can be
determined from Eq. (\ref{pert4}) by measuring $f$, $f_0$, and calculating
$|H_{\rm max}|^2/\int_{V_c}|H_0|^2 dv$ for given cavity dimensions and known
magnetic field distribution.

\noindent
For example, taking a rectangular prism-like sample of the same aspect ratio
as a hypothetic ellipsoidal sample we find
\begin{equation}
\label{pert6}
\frac{f-f_0}{f_0}=\frac{1}{1-N_{\rm eff}}\cdot
\frac{abc|H_{\rm max}|^2}{2\int_{V_c}|H_0|^2 dv},
\end{equation}
where $1-N_{\rm eff}=\alpha\cdot(c/b){\rm\bf{E}}(k)$
and $\alpha$ is a parameter taking into account the field distortions due to
sharp edges and corners of a rectangular sample.

Substituting $1-N_{\rm eff}$ into Eq. (\ref{pert6}) we obtain finally
\begin{equation}
\label{pert7}
\frac{f-f_0}{f_0}=\frac{ab^2}
{\alpha{\rm\bf{E}}(k)}\cdot
\frac{|H_{\rm max}|^2}{2\int_{V_c}|H_0|^2 dv},
\end{equation}
where $k=[1-(b/a)^2]^{1/2}$, and in the limiting case $T\rightarrow 0$
 Eq. (\ref{pert7}) can be used to define a geometrical factor
$\Gamma_\bot$:
\begin{equation}
\label{pert8}
\frac{f(0)-f_0(0)}{f_0(0)}=\frac{ab^2}
{\alpha{\rm\bf{E}}(k)}\cdot
\frac{|H_{\rm max}|^2}{2\int_{V_c}|H_0|^2 dv}=\Gamma_\bot,
\end{equation}
where $f(0)$, $f_0(0)$ denote the resonant frequencies of the perturbed
and unperturbed cavity extrapolated to $T=0$.

\noindent
It should be mentioned here that the detailed information about the parameter
$\alpha$ is not required at this point, since Eq. (\ref{pert8}) defines
already an {\it effective} value of $\Gamma_\bot$ taking into account
field distortions around a rectangular sample.

\noindent
Note that in the perpendicular orientation the current flows only in the
$ab$ plane, thus the corresponding penetration depth will be denoted as
$\lambda_{ab}$. Taking $a-2\lambda_{ab}$, $b-2\lambda_{ab}$ instead
of $a$, $b$ we find
\begin{equation}
\label{lambda1}
(a-2\lambda_{ab})(b-2\lambda_{ab})^2=ab^2\left(1-\frac{2\lambda_{ab}}{a}\right)
\left(1-\frac{2\lambda_{ab}}{b}\right)^2
\simeq ab^2\left[1-2\lambda_{ab}\left(\frac{1}{a}+\frac{2}{b}\right)\right].
\end{equation}

For $\lambda_{ab}>0$ we find from Eq. (\ref{lambda1})
\begin{equation}
\label{pert9}
\frac{f(T)-f_0(T)}{f_0(T)}=\Gamma_\bot
\left[1-2\lambda_{ab}(T)\left(\frac{1}{a}+\frac{2}{b}\right)\right].
\end{equation}

\noindent
Subtracting both sides of Eqs. (\ref{pert8}) and (\ref{pert9}),
and using $f_0(T)\simeq f_0(0)$ in the denominator we find finally
\begin{equation}
\label{lambda2}
\frac{\delta f(T)}{f_0(0)}=\Gamma_\bot\frac{2\lambda_{ab}(T)}{a}
\left(1+2\frac{a}{b}\right),
\end{equation}
thus
\begin{equation}
\label{lambda3}
\lambda_{ab}(T)=\frac{\delta f(T)}{f_0(0)}\cdot\frac{a}{2\Gamma_\bot
(1+2a/b)},
\end{equation}
where $\delta f(T)=[f(0)-f(T)]-[f_0(0)-f_0(T)]$.

\noindent
Strictly speaking, both $\alpha$ and ${\rm\bf{E}}(k)$ in Eq. (\ref{pert7})
are weakly dependent on the sample dimensions $a$ and $b$. More detailed
analysis shows, however, that the corrections due to $\alpha$ and
${\rm\bf{E}}(k)$ are of opposite sign and are at least one order of
magnitude smaller than the leading term in Eq. (\ref{lambda1}).
Therefore the influence of $\alpha(a,b)$ and ${\rm\bf{E}}(a,b)$ on the
final result has been neglected in the above analysis.

\subsubsection{Parallel orientation}

\noindent
Contrary to the perpendicular orientation, now the field distortion
is very small and for a very flat
ellipsoid ($\bar{a}>\bar{b}\gg\bar{c}$) we have
\begin{equation}
\label{n1}
N\simeq \frac{\pi\bar{c}}{4\bar{a}}\ll 1,
\end{equation}
where the magnetic field is parallel to the flat face $ab$ of the sample.

\noindent
For a rectangular prism ($a>b\gg c$) we can assume also $N\propto c/a\ll 1$,
thus $N$ in Eq. (\ref{pert4}) can be neglected with respect to unity.
Substituting the sample volume $V_s=abc$ and assuming $\lambda(0)=0$
we can use Eq. (\ref{pert4}) to define the geometric factor of the sample as
\begin{equation}
\label{pert2a}
\frac{f(0)-f_0}{f_0}=\frac{|H_{\rm max}|^2 abc}{2\int H^2dv}=\Gamma_\|,
\end{equation}
where $f(0)$ denotes the resonant frequency of the cavity with a sample,
extrapolated to $T=0$, and $\Gamma_\|$ is the geometric factor in the
parallel orientation.

\noindent
For $T>0$ the magnetic field penetrates into the sample from both sides,
thus
\begin{equation}
\label{abc}
abc\rightarrow a(b-2\lambda_c)(c-2\lambda_{ab})
\simeq abc\left[1-2\left(\frac{\lambda_c}{b}+
\frac{\lambda_{ab}}{c}\right)\right],
\end{equation}
and substituting Eq. (\ref{abc}) into Eq. (\ref{pert2a}) we obtain
\begin{equation}
\label{pert3a}
\frac{f(T)-f_0}{f_0}=\Gamma_\|\left[1-2\left(
\frac{\lambda_c}{b}+\frac{\lambda_{ab}}{c}\right)\right],
\end{equation}
where $f(T)$ denotes the resonant frequency of  the cavity containing
a sample, measured at $T>0$.\\

\noindent
Subtracting Eqs. (\ref{pert3a}) and (\ref{pert2a}) we find
\begin{equation}
\label{pert4a}
\frac{\delta f(T)}{f_0}=2\Gamma_\|\left(
\frac{\lambda_c}{b}+\frac{\lambda_{ab}}{c}\right)=
\frac{|H_{\rm max}|^2}{\int |H_0|^2dv}(ab\lambda_{ab}+ac\lambda_c),
\end{equation}
where $\delta f(T)=[f(0)-f(T)]-[f_0(0)-f_0(T)]$, $\lambda_{ab}$ and $\lambda_c$
denote the anisotropic penetration depth for a superconducting sample.\\

\noindent
Summarizing, for given frequency shift $\delta f(T)$ and the factor
$|H_{\rm max}|^2/\int |H_0|^2dv$, Eq. (\ref{pert4a}) makes it possible
to determine a linear combination of the anisotropic penetration depth
$\lambda_{ab}$ and $\lambda_c$. Next, we can calculate
$\lambda_c$, provided $\lambda_{ab}$ is known (e.g. from earlier
measurements in the perpendicular orientation).

\end{appendix}

\end{document}